\documentclass[12pt]{revtex4}
\usepackage{dcolumn}
\usepackage{graphicx}
\usepackage{amsmath}
\usepackage{amsfonts}
\usepackage{amssymb}
\usepackage{psfrag}
\usepackage{wrapfig}
\usepackage{subfigure}
\usepackage{makeidx}
\usepackage{bm}
\usepackage{epsf}
\usepackage{hyperref}

\begin{document}

\title{A new metric for rotating charged Gauss-Bonnet black holes in AdS spaces}

\author{ De-Cheng Zou$^{1}$\footnote{ Email:zoudecheng789@163.com}, Zhan-Ying Yang$^{1}$\footnote{ Email:zyyang@nwu.edu.cn},
Rui-Hong Yue$^{2}$\footnote{Email:yueruihong@nbu.edu.cn} and Tian-Yi Yu$^{2}$}

\affiliation{ $^{1}$Department of Physics, Northwest University, Xi'an, 710069, China\\
$^{2}$Faculty of Science, Ningbo University, Ningbo, 315211, China}
\date{\today}

\begin{abstract}
\indent

This paper presents a new metric for slowly rotating charged Gauss-Bonnet
black holes in higher dimensional anti-de Sitter spaces. Taking the angular
momentum parameter $a$ up to second order, the slowly rotating charged black hole
solutions are obtained by working directly in the action.
\end{abstract}



\maketitle

\section{Introduction}  
\label{11s}
The higher derivative gravity theories have gained much interest during the past years.
It is believed that Einstein's action is only an effective gravitational action valid
for small curvatures or low energies \cite{Myers:1988ze}. Considering the fundamental
nature of quantum gravity, one sees that the action will be modified by higher derivative
interactions in the the renormalization of quantum field theory in curved
spacetimes \cite{Stelle:1976gc, deBerredoPeixoto:2003pj, deBerredoPeixoto:2004if}.
Besides, these higher derivative terms can be seen in
the construction of the low-energy effective action of
string \cite{Lust:1989tj, Myers:1987yn, Alekseev:1997wy, Maeda:2009uy}.
The so-called Lovelock gravity is quite special \cite{Lovelock:1971yv}.
Except for the advantage that the equations of motion of the Lovelock gravity,
as the case of the Einstein's general relativity, do not contain terms with
more than second derivatives of metric, the Lovelock gravity has been shown
to be free of ghost when expanding on a flat space, evading any problems with
unitarity \cite{Zwiebach:1985uq, Zwiebach:1986uq}. The Gauss-Bonnet terms
appear as the first higher derivative curvature correction term to general relativity.
The black hole solutions in Gauss-Bonnet gravity were first discovered by Boulware
and Deser \cite{Boulware:1985wk} and Wheeler \cite{Wheeler:1985qd}, independently.
The thermodynamics of the uncharged static spherically Gauss-Bonnet black hole solutions
have been considered in Refs \cite{Cai:2003gr, Cai:2001dz, Cho:2002hq}
and of charged solutions in Ref \cite{Cvetic:2001bk}.
Recently the quasinormal mode of a scalar field in five dimensional
Lovelock black hole spacetime for different angular quantum numbers $l$ has been
obtained in \cite{Chen:2009an}. Liu \cite{Liu:2007zze} studied the electromagnetic
perturbations of black holes in Gauss-Bonnet gravity.

It is of interest to generalize these static Gauss-Bonnet solutions by including the
effects of rotation. This problem has been considered recently
in \cite{Zou:2010dx} within a perturbative approach.
It is worth mentioning that, for the Gauss-Bonnet action,
the resulting field equations, obtained after variation with respect
to the metric tensor, have seven terms. While the resulting field
equation for third order Lovelock gravity contains thirty-four terms.
However, by working directly in the action, the exact static Gauss-Bonnet black hole
solutions were obtained \cite{Boulware:1985wk, Wheeler:1985qd, Cai:2003gr, Cai:2001dz, Cho:2002hq},
and third order Lovelock black hole solutions in \cite{Dehghani:2009zzb, Ge:2009ac, Zou}.
Using the same approach, in this paper, we want to obtain the slowly
rotating black hole solutions in Lovelock gravity.
Apparently the lowest level contribution of rotation
should be proportional to $a^2$. Hence, linearly dependent on $a$,
the metric demonstrated in Ref \cite{Zou:2010dx} can not be adopted in this case.
So, we need to find a proper metric ansatz up to $a^2$.

This paper is organized as follows. In Section 2, we present a new form metric and
obtain slowly rotating charged black hole solutions by working directly in the action.
Then we discuss some related thermodynamic properties of black holes.
Section 3 is devoted to a summary of the results.

\section{Slowly rotating charged Gauss-Bonnet black holes}

\subsection{A new metric and Rotating black holes}
\indent

In general relativity, the higher dimensional generalization of the Kerr-AdS
black hole solutions can be found in \cite{Aliev:2007qi, Aliev:2006yk}.
It is expressed as in the Boyer-Lindpuist type coordinate
\begin{eqnarray}
ds^2=&-&\frac{\Delta_r}{\Sigma}(dt-\frac{a\sin^2\theta}{\Xi}d\phi)^2+\frac{\Sigma}{\Delta_r}dr^2
+\frac{\Sigma}{\Delta_{\theta}}d\theta^2+\frac{\Delta_{\theta}\sin^2\theta}{\Sigma}(a dt
-\frac{r^2+a^2}{\Xi}d\phi)^2\nonumber\\
&+&r^2\cos^2\theta d\Omega^2_{D-4},\label{eq:1b}
\end{eqnarray}
where
\begin{eqnarray}
\Sigma&=&r^2+a^2\cos^2\theta,\quad \Delta_r=(r^2+a^2)(1+\frac{r^2}{l^2})-mr^{5-D},\nonumber\\
\Delta_{\theta}&=&1-\frac{a^2}{l^2}\cos^2\theta,\qquad \Xi=1-\frac{a^2}{l^2}.\nonumber
\end{eqnarray}
The parameters $m$ and $a$ are related to the mass and angular
momentum of the black hole, as we shall see.

If we consider the metric Eq.~(\ref{eq:1b}) up to the second order in the parameter $a$,
the components of the metric are expressed as follows:
\begin{eqnarray}
g_{tt}&=&-[1-\frac{m}{r^{D-3}}+\frac{r^2}{l^2}+\frac{a^2}{l^2}\sin^2\theta
+\frac{a^2m\cos^2\theta}{r^{D-1}}]+\mathcal {O}(a^4),\nonumber\\
g_{rr}&=&\frac{1+a^2\cos^2\theta/r^2}{1-\frac{m}{r^{D-3}}+\frac{r^2}{l^2}}
-\frac{a^2}{(1-\frac{m}{r^{D-3}}+\frac{r^2}{l^2})^2}(\frac{1}{l^2}
+\frac{1}{r^2})+\mathcal {O}(a^4),\nonumber\\
g_{\theta\theta}&=&(r^2+a^2\cos^2\theta+\frac{a^2r^2}{l^2}\cos^2\theta)
+\mathcal {O}(a^4),\nonumber\\
g_{\phi\phi}&=&[r^2+a^2+a^2\sin^2\theta(\frac{m}{r^{D-3}}-\frac{r^2}{l^2})
+\frac{2a^2r^2\sin^2\theta}{l^2}]\sin^2\theta+\mathcal {O}(a^4),\nonumber\\
g_{t\phi}&=&2a\sin^2\theta(\frac{r^2}{l^2}-\frac{m}{r^{D-3}})+\mathcal {O}(a^3).\label{eq:2b}
\end{eqnarray}
Setting $a=0$, we find that metric Eq.~(\ref{eq:2b}) reduce to the static case.
It is known that the static black hole solution in D dimensional spacetimes
in general relativity is $f(r)=1-\frac{m}{r^{D-3}}+\frac{r^2}{l^2}$.
Replacing $1-\frac{m}{r^{D-3}}+\frac{r^2}{l^2}$ with $f(r)$, we obtain a new metric
when all terms involving $a^3$ and higher powers in $a$ are ignored
\begin{eqnarray}
ds^2=&-&\Big[f(r)+\frac{a^2}{l^2}+\frac{a^2(1-f(r))\cos^2\theta}{r^2}\Big]dt^2\nonumber\\
&+&\Big[\frac{1}{f(r)}+\frac{a^2(\cos^2\theta f(r)-1)}{r^2f(r)^2}
-\frac{a^2}{l^2f(r)^2}\Big]dr^2\nonumber\\
&+&(r^2+a^2\cos^2\theta+\frac{a^2r^2}{l^2}\cos^2\theta)d\theta^2
+\Big[r^2+a^2\nonumber\\
&+&a^2\big(1-f(r)\big)\sin^2\theta+\frac{2a^2r^2\sin^2\theta}{l^2}\Big]\sin^2\theta d\phi^2\nonumber\\
&+&2ar^2 p(r)\sin^2\theta dtd\phi+r^2\cos^2\theta d\Omega_{D-4}^2,\label{eq:1a}
\end{eqnarray}
where parameter $a^2$ is a small quantity and the functions $f(r)$ and $p(r)$ satisfy the relation
$p(r)=\frac{-1+f(r)}{r^2}$. Simplification for future, we order $f(r)$ and $p(r)$
are two independent variables. Besides, taking the parameter $a$ up to the first order,
the metric Eq.~(\ref{eq:1a}) becomes the counterpart in Ref \cite{Zou:2010dx}.

The action for Gauss-Bonnet theory with negative cosmological constant $\Lambda=-(D-1)(D-2)/l^2$
in D dimensions is given by \cite{Zou:2010dx}
\begin{eqnarray}
{\cal I}=\frac{1}{16\pi G}\int d^{D}x\sqrt{-g}(-2\Lambda+R+\alpha {\cal L}_{GB}
-4\pi G F_{\mu\nu}F^{\mu\nu}),\label{eq:2a}
\end{eqnarray}
where $\alpha$ is the Gauss-Bonnet coefficient with dimension $(length)^2$ and
is rescaled with $\frac{\tilde{\alpha}}{(D-3)(D-4)}$ in future.
Note that the Lagrangian of Lovelock gravity is the sum of dimensionally extended Euler
densities, and then, the Lagrangian of Gauss-Bonnet term also can be expressed as \cite{Cai:2001dz}
\begin{eqnarray}
{\cal L}_{GB}=\frac{1}{4}\delta^{a_1b_1a_2b_2}_{c_1b_1c_2d_2}
R^{c_1d_1}_{~~~~a_1b_1}R^{c_2d_2}_{~~~~a_2b_2}.\label{eq:3a}
\end{eqnarray}

Here, all non-vanishing components of Riemann tensors (up to $a^2$) can be expressed as
\begin{eqnarray}
R^{ij}_{~~kl}=\bar{R}^{ij}_{~~kl}+\tilde{R}^{ij}_{~~kl},\nonumber
\end{eqnarray}
where $\bar{R}^{ij}_{~~kl}$ is in the absence of $a$ and $\tilde{R}^{ij}_{~~kl}$
is proportional to $a^2$. Moreover, based on the relationship of indexes,
these Riemann tensors also can be classified into three groups:
(I) diagonal form $R^{\hat{i}\hat{j}}_{~~\hat{i}\hat{j}}$, (II) off-diagonal form
proportional to $a$ and (III) off-diagonal form proportional to $a^2$.
For group (I), we find that the Riemann tensors $\bar{R}^{12}_{~~12}$, $\bar{R}^{1i}_{~~1i}$,
$\bar{R}^{2i}_{~~2i}$ and $\bar{R}^{ij}_{~~ij}$ are all absent from parameter $a$.
They can be represented as $\bar{R}^{12}_{~~12}=-\frac{f''(r)}{2}$,
$\bar{R}^{1i}_{~~1i}=-\frac{f'(r)}{2r}$, $\bar{R}^{2i}_{~~2i}=-\frac{f(r)}{2r}$
and $\bar{R}^{ij}_{~~ij}=\frac{1-f(r)}{r^2} (3\leq i<j \leq D)$
which is independent to the dimensions $D$.
Some of key steps are given in appendix. Denoted the Lagrangian of Gauss-Bonnet term
from the contribution of the case (I) by ${\cal L}_d$, it can be written as
\begin{eqnarray}
{\cal L}_d&=&{\cal \bar{L}}_d+{\cal \tilde{L}}_d\nonumber\\
&=&\frac{1}{4}\delta^{a_1b_1a_2b_2}_{c_1b_1c_2d_2}(\bar{R}^{c_1d_1}_{
~~~~a_1b_1}\bar{R}^{c_2d_2}_{~~~~a_2b_2}
+2\bar{R}^{c_1d_1}_{~~~~a_1b_1}\tilde{R}^{c_2d_2}_{~~~~a_2b_2})\nonumber\\
&=&\bar{A}_0+\bar{A}_1+\bar{A}_2+\bar{A}_3+\bar{A}_4+\tilde{B}_1+\tilde{B}_2
+\tilde{B}_3+\tilde{B}_4,\label{eq:4a}
\end{eqnarray}
where
\begin{eqnarray}
\bar{A}_0&=&\frac{1}{4}\delta^{a_1b_1 a_2b_2}_{c_1b_1 c_2d_2}\bar{R}^{c_1d_1}_{~
~~~a_1b_1}\bar{R}^{c_2d_2}_{~~a_2b_2}\nonumber\\
&=&(D-2)(D-3)(D-4)(D-5)(\bar{R}^{ij}_{~~ij})^2,\quad (i,j\geq 3)\nonumber\\
\bar{A}_1&=&2\delta^{a_1b_1 12}_{c_1b_1 12}\bar{R}^{c_1d_1}_{~~~~a_1b_1}\bar{R}^{12}_{~~12}
=4(D-2)(D-3)\bar{R}^{12}_{~~12}\bar{R}^{ij}_{~~ij},\nonumber\\
\bar{A}_2&=&\frac{1}{4}\delta^{1i2j}_{1i2j}\bar{R}^{1i}_{~~1i}\bar{R}^{2j}_{~~2j}
=8(D-2)(D-3)\bar{R}^{1i}_{~~1i}\bar{R}^{2j}_{~~2j},\nonumber\\
\bar{A}_3&=&\frac{1}{4}\delta^{1ijk}_{1ijk}\bar{R}^{jk}_{~~jk}\bar{R}^{1i}_{~~1i}
=4(D-2)(D-3)(D-4)\bar{R}^{1i}_{~~1i}\bar{R}^{jk}_{~~jk},\nonumber\\
\bar{A}_4&=&\frac{1}{4}\delta^{2ijk}_{2ijk}\bar{R}^{jk}_{~~jk}\bar{R}^{2i}_{~~2i}
=4(D-2)(D-3)(D-4)\bar{R}^{2i}_{~~2i}\bar{R}^{jk}_{~~jk},\quad (i,j,k\geq 3)\label{eq:5a}
\end{eqnarray}
and
\begin{eqnarray}
\tilde{B}_1&=&4(D-2)(D-3)\tilde{R}^{12}_{~~12}\times\bar{R}^{34}_{~~34},\nonumber\\
\tilde{B}_2&=&4\sum^{D}_{i=3}\tilde{R}^{1i}_{~~1i}\times\Big[2(D-3)\bar{R}^{24}_{~~24}
+(D-3)(D-4)\bar{R}^{45}_{~~45}\Big],\nonumber\\
\tilde{B}_3&=&4\sum^{D}_{i=3}\tilde{R}^{2i}_{~~2i}\times\Big[2(D-3)\bar{R}^{14}_{~~14}
+(D-3)(D-4)\bar{R}^{45}_{~~45}\Big],\nonumber\\
\tilde{B}_4&=&8\sum^{D}_{i>j=3}\tilde{R}^{ij}_{~~ij}\times\Big[\bar{R}^{12}_{~~12}
+(D-4)(\bar{R}^{13}_{~~13}+\bar{R}^{25}_{~~25})\nonumber\\
&+&\frac{(D-4)(D-5)}{2}\bar{R}^{kl}_{~~kl}\Big], \quad (k>l \geq 4) .\label{eq:6a}
\end{eqnarray}

Inspecting the off-diagonal Riemann tensors, we calculate the parts of Lagrangian
from the off-diagonal Riemann tensors in case (II)
\begin{eqnarray}
{\cal L}_{od}={\cal L}_{od1}+{\cal L}_{od2}+{\cal L}_{od3},\label{eq:7a}
\end{eqnarray}
where
\begin{eqnarray}
{\cal L}_{od1}&=&4(R^{12}_{~~34}R^{34}_{~~12}+R^{13}_{~~24}R^{24}_{~~13}
+R^{14}_{~~23}R^{23}_{~~14}),\nonumber\\
{\cal L}_{od2}&=&\sum_{i=3}^{D}4(R^{12}_{~~24}R^{i4}_{~~1i}
+R^{24}_{~~12}R^{1i}_{~~i4}),\nonumber\\
{\cal L}_{od3}&=&\sum_{i,j=3}^{D}4(R^{1i}_{~~i4}R^{j4}_{~~1j}
+R^{1j}_{~~j4}R^{i4}_{~~1i}),\quad (i>j \geq 3).\label{eq:8a}
\end{eqnarray}
It is interesting to note that the contribution of off-diagonal Riemann tensors in
case (III) vanishes following the properties of the Kronecker delta symbol.
In addition, the Ricci scalar $R$ is given by
\begin{eqnarray}
R=\bar{R}+\tilde{R},\label{eq:9a}
\end{eqnarray}
where $\bar{R}$ is equal to $-\frac{6f'(r)}{r}+\frac{6(1-f(r))}{r^2}-f''(r)$ and $\tilde{R}$
is proportional to $a^2$ and shown in appendix.

Since the black hole rotates along the direction $\phi$,
it will generate a magnetic field.
If we considering this effect, the gauge potential can be chosen
\begin{eqnarray}
A_{\mu}dx^{\mu}=A_{t}dt+A_{\phi}d\phi,\label{eq:10a}
\end{eqnarray}
where $A_t=Qh(r)+a^2\cos^2\theta Q k(r),\quad A_{\phi}=-a\sin^2\theta Q c(r)$.
Therefore the nonzore contravariant component of the electromagnetic field
tensor can be written as
\begin{eqnarray}
F_{tr}&=&-Q h'(r)-a^2\cos^2\theta Q k'(r),\quad F_{t\theta}=a^2\sin(2\theta)Q k(r)\nonumber\\
F_{r\phi}&=&-a Q c'(r)\sin^2\theta, \quad F_{\theta\phi}=-a Q c(r)\sin(2\theta).\label{eq:11a}
\end{eqnarray}
By substituting the ansatz Eq.~(\ref{eq:1a}) and tensors of electromagnetic
field Eq.~(\ref{eq:11a}) into the action Eq.~(\ref{eq:2a}), the action becomes
\begin{eqnarray}
{\cal I}={\cal I}(f(r),p(r), h(r), k(r), c(r)).\label{eq:12a}
\end{eqnarray}

Firstly, varying the action Eq.~(\ref{eq:12a}) with regard to the function $k(r)$,
we find the solution
\begin{eqnarray}
h''(r)r+(D-2)h'(r)=0.\label{eq:13a}
\end{eqnarray}
Hence, we obtain
\begin{eqnarray}
h(r)=\frac{C_2}{r^{D-3}}+C_1.\label{eq:14a}
\end{eqnarray}
where $C_1$ and $C_2$ are two integration constants. We choose $C_1=0$
and $C_2=\frac{1}{4(D-3)\pi}$
and then the function $h(r)$ can be written as $h(r)=\frac{1}{4\pi(D-3)r^{D-3}}$.
Then, the variation of the action Eq.~(\ref{eq:12a}) with regard to $c(r)$ becomes
\begin{eqnarray}
[r^{D-4}f(r)c'(r)]'-2(D-3)r^{D-6}c(r)+\frac{p'(r)}{4\pi}=0.\label{eq:15a}
\end{eqnarray}
Taking the variation of the action Eq.~(\ref{eq:12a}) with regard to the function $p(r)$,
we arrive at
\begin{eqnarray}
0&=&[2\tilde{\alpha} f(r)-2\tilde{\alpha}-r^2]r^2f(r)p''(r)\nonumber\\
&+&[2(D-2)\tilde{\alpha}(f(r)-1)-Dr^2+2\tilde{\alpha} r f'(r)]r f(r)p'(r)\nonumber\\
&+&\{(D-2)[-f'(r) r^3+(1-f(r))((D-3)r^2-2\tilde{\alpha} r f'(r))+(D-5)\tilde{\alpha} (1-f(r))^2\nonumber\\
&+&\frac{(D-1)r^4}{l^2}]-8\pi G Q^2 r^4h'(r)\}p(r)+16\pi G Q^2 r h'(r)c'(r).\label{eq:16a}
\end{eqnarray}
If supposing the coefficient of function $p(r)$ vanishes, we find that,
for $h(r)=\frac{1}{4\pi(D-3)r^{D-3}}$, the function $f(r)$ is obtained
\begin{eqnarray}
f(r)=1+\frac{r^2}{2\tilde{\alpha}}(1-\sqrt{1-\frac{4\tilde{\alpha}}{l^2}
+\frac{4\tilde{\alpha} m}{r^{D-1}}-\frac{2\tilde{\alpha}}{\pi (D-2)(D-3)}\frac{GQ^2}{r^{2D-4}}}).\label{eq:17a}
\end{eqnarray}
Moreover, the Eq.(\ref{eq:16a}) reduces to
\begin{eqnarray}
r^{D}[1+2\tilde{\alpha}\omega(r)]p'(r)+4GQ^2c(r)+C_3=0,\label{eq:18a}
\end{eqnarray}
where $\omega(r)=(1-f(r))/r^2$. Based on the Ref.\cite{Zou:2010dx},
the function $p(r)$ is given by
\begin{eqnarray}
p(r)=\frac{1}{2\tilde{\alpha}}(1-\sqrt{1-\frac{4\tilde{\alpha}}{l^2}+\frac{4\tilde{\alpha} m}{r^{D-1}}
-\frac{2\tilde{\alpha}}{\pi (D-2)(D-3)}\frac{GQ^2}{r^{2D-4}}}).\label{eq:19a}
\end{eqnarray}
and $c(r)$ is
\begin{eqnarray}
c(r)=-\frac{1}{4(D-3)\pi r^{D-3}},\label{eq:20a}
\end{eqnarray}
by taking constant $C_3=(D-1)m$. Apparently the functions $f(r)$ and $p(r)$ satisfy
the relationship $p(r)=-(1-f(r))/r^2$.

Furthermore, varying the action with regard to the electromagnetic field $h(r)$, we have
\begin{eqnarray}
(D-1)+2(D-2)(D-3)\pi r^{D}k'(r)+2(D-3)\pi r^{D+1}k''(r)=0.\label{eq:22a}
\end{eqnarray}
We can obtain $k(r)=-\frac{1}{4\pi(D-3)r^{D-1}}$ where integral constants are all ignored.
Therefore, the tensor of electromagnetic field are given as
\begin{eqnarray}
F_{tr}&=&\frac{Q}{4\pi r^{D-2}}-\frac{(D-1)}{4\pi(D-3) r^D}a^2Q\cos^2\theta,\quad
F_{t\theta}=-\frac{a^2Q}{4\pi(D-3) r^{D-1}}\sin(2\theta)\nonumber\\
F_{r\phi}&=&-\frac{a Q}{4\pi r^{D-2}}\sin^2\theta,\quad
F_{\theta\phi}=\frac{a Q}{4\pi(D-3)r^{D-3}}\sin(2\theta).\label{eq:23a}
\end{eqnarray}
Furthermore, it is easy to verify that the expressions for functions $f(r)$ Eq.~(\ref{eq:18a}),
$p(r)$ Eq.~(\ref{eq:19a}), $c(r)$ Eq.~(\ref{eq:20a}) and $h(r)=-k(r)r^2=\frac{1}{4\pi(D-3)r^{D-3}}$
still satisfy the variation of action Eq.~(\ref{eq:2a}) respecting to $f(r)$.

\subsection{Thermodynamics of black holes}

The Killing vectors can be used to give a physical interpretation of the parameter
$m$ and $a$. Following the analysis given in \cite{Aliev:2006yk, Komar:1958wp, Peng:2007pk, Zeng:2008zza},
one can obtain coordinate-independent definitions for these parameters. We have the integral
\begin{eqnarray}
M=-\frac{1}{16\pi G}\frac{D-2}{D-3}\oint \xi^{\mu;\nu}_{(t)}d^{D-2}\Sigma_{\mu\nu},\quad
J=\frac{1}{16\pi G}\oint \xi^{\mu;\nu}_{(\phi)}d^{D-2}\Sigma_{\mu\nu},\label{eq:24a}
\end{eqnarray}
where the integrals are taken over the $(D-2)$-sphere at spatial infinity,
\begin{eqnarray}
d^{D-2}\Sigma_{\mu\nu}=\frac{1}{(D-2)!}\sqrt{-g}\epsilon_{\mu\nu\alpha\beta\gamma}dx^{i_1}\wedge
dx^{i_2}\wedge...dx^{i_{D-2}}.\label{eq:25a}
\end{eqnarray}
To justify the definitions Eq.~(\ref{eq:24a}), we can calculate the integrands in the asymptotic
region $r\rightarrow \infty$. For the dominant terms in the asymptotic expansion we have
\begin{eqnarray}
\xi^{t;r}_{(t)}=\frac{m(D-3)}{2r^{D-2}}+\mathcal {O}(\frac{1}{r^{D}}),\quad
\xi^{t;r}_{(\phi)}=-\frac{am(D-1)\sin^2\theta}{2r^{D-2}}+\mathcal {O}(\frac{1}{r^{D}}).\label{eq:26a}
\end{eqnarray}
We perform the integration over a $(D-2)$-sphere at $r \rightarrow \infty$. It gives
\begin{eqnarray}
M=\frac{m(D-2)\Sigma_k}{16\pi G},\quad J=\frac{2Ma}{D-2}.\label{eq:27a}
\end{eqnarray}

In addition, if $a=0$, the black hole does not rotate and the metric represents a spherically
symmetric charged black hole with a spherically symmetric electric field.
If the black hole rotates $a \neq 0$, the electric field is supplemented by a magnetic
field due to the dragging of the inertial reference frames into rotational motion around
the black hole.

\section{Concluding remarks}

In present paper, we proposed an new metric and then obtained
the slowly rotating charged Gauss-Bonnet black hole
solutions in anti-de Sitter spaces by working directly in the action.
Besides the function $f(r)$, the diagonal components of the metric also
involve $a^2$. Moreover, the $g_{t\phi}$ is proportional to $r^2p(r)$ as to make the
equation for $p(r)$ much simple. Since the black
hole rotates along the direction $\phi$, vector potential has an extra nonradial
component $A_{\phi}=-a\sin^2\theta Q c(r)$ and the $t$ component is
written as $A_t=Qh(r)+a^2\cos^2\theta Q k(r)$.
By taking variation of the action respecting to
the functions $p(r)$, $f(r)$, $c(r)$, $h(r)$ and $k(r)$, respectively,
we got the exact form for function $p(r)$,
while the function $f(r)$ still kept the form of the static solutions. Moreover,
the exact expressions for $c(r)$, $h(r)$ and $k(r)$ have been expressed as
$c(r)=-h(r)=k(r)r^2=-\frac{1}{4\pi(D-3)r^{D-3}}$.

\section{Appendix}
\indent
From the metric Eq.~(\ref{eq:1a}), we obtain some of the intermediate steps of the calculation.

\textbf{Riemann tensors}. The non-vanishing Riemann tensors can be classified into three
groups: (I) diagonal form $R^{\hat{i}\hat{j}}_{~~\hat{i}\hat{j}}$, (II) off-diagonal form
proportional to $a$ and (III) off-diagonal form proportional to $a^2$.

group (I):
\begin{eqnarray}
R^{12}_{~~12}&=&\bar{R}^{12}_{~~12}+\tilde{R}^{12}_{~~12},\quad
\bar{R}^{12}_{~~12}=-\frac{f(r)''}{2},\nonumber\\
\tilde{R}^{12}_{~~12}&=&a^2\left\{[(-1+\frac{r f(r)'+r^2 f(r)''}{2f(r)}
-\frac{r^2f(r)'^2}{4f(r)^2})p(r)^2\right.\nonumber\\
&+&\left.(\frac{r^2f(r)'p(r)'}{f(r)}-5rp(r)'-r^2p(r)'')\frac{p(r)}{2}
-\frac{r^2p(r)'^2}{4}+\frac{f(r)'-r f(r)''}{2r^3f(r)}\right.\nonumber\\
&+&\left.\frac{f(r)'^2}{4r^2f(r)^2}]\sin^2\theta+[\frac{-2f(r)'}{r^3}
-\frac{3(1-f(r))}{r^4}+\frac{f(r)''}{2r^2}]\cos^2\theta\right\},\nonumber
\end{eqnarray}

\begin{eqnarray}
R^{1i}_{~~1i}&=&\bar{R}^{1i}_{~~1i}+\tilde{R}^{1i}_{~~1i},\quad
\bar{R}^{1i}_{~~1i}=-\frac{f(r)'}{2r},\quad (3\leq i\leq D) ,\nonumber\\
\tilde{R}^{13}_{~~13}&=&\left\{-\frac{p(r)^2}{f(r)}+\frac{1}{f(r)r^4}+\frac{f(r)'}{r^3}-\frac{f(r)}{r^4}
+[-\frac{r p(r)^2f(r)'}{2f(r)}+\frac{r p(r)p(r)'}{2}\right.\nonumber\\
&-&\left.\frac{1+f(r)}{r^4}-\frac{f(r)'}{2f(r)r^3}+\frac{2}{r^4f(r)}-\frac{2p(r)^2}{f(r)}
+\frac{f(r)'}{r^3}+p(r)^2]\sin^2\theta\right\}a^2\nonumber\\
\tilde{R}^{14}_{~~14}&=&\left\{[-p(r)^2-r p(r)'p(r)-\frac{r^2p(r)'^2}{4}+\frac{f(r)'^2}{r^4}
-\frac{f(r)'}{2r^3f(r)}+\frac{f(r)'}{4r^2}\right.\nonumber\\
&-&\left.\frac{f(r)'f(r)}{2r^3}+\frac{r f(r)'p(r)^2}{2f(r)}]\sin^2\theta+\frac{f(r)'}{r^3}
+\frac{-p(r)^2r^4+1-f(r)^2}{f(r)r^4}\cos^2\theta\right\}a^2\nonumber\\
\tilde{R}^{1j}_{~~1j}&=&\left\{[\frac{p(r)^2}{f(r)}-p(r)^2-\frac{f(r)'}{2r^3f(r)'}
-\frac{1}{r^4f(r)}+\frac{f(r)'p(r)^2}{2f(r)}-\frac{r p(r)'p(r)}{2}]\sin^2\theta\right. \nonumber\\
&+&\left. [\frac{f(r)'}{2r^3}-\frac{f(r)}{r^4}]\cos^2\theta+\frac{1}{r^4}\right\}a^2,\quad (5\leq j \leq D),\nonumber
\end{eqnarray}

\begin{eqnarray}
R^{2i}_{~~2i}&=&\bar{R}^{2i}_{~~2i}+\tilde{R}^{2i}_{~~2i},\quad
\bar{R}^{2i}_{~~2i}=-\frac{f(r)}{2r}, \quad (3\leq i \leq D),\nonumber\\
\tilde{R}^{23}_{~~23}&=&\tilde{R}^{2j}_{~~2j}=a^2[\frac{f(r)'}{r^3}
+\frac{2(1-f(r))}{r^4}]\cos^2\theta, \quad (5\leq j \leq D),\nonumber\\
\tilde{R}^{24}_{~~24}&=&\left\{[\frac{f(r)'^2-2f(r)''f(r)}{4r^2}-1-\frac{r^2p(r)p(r)''}{2}
-\frac{3f(r)f(r)'}{2r^3}+\frac{f(r)^2}{r^4}\right.\nonumber\\
&-&\left.\frac{r^2p(r)'^2}{4}-\frac{3rp(r)p(r)'}{2}]\sin^2\theta
+4(2-2f(r)+f(r)'r)\right\}a^2,\nonumber
\end{eqnarray}

\begin{eqnarray}
R^{ij}_{~~ij}&=&\bar{R}^{ij}_{~~ij}+\tilde{R}^{ij}_{~~ij},\quad
\bar{R}^{ij}_{~~ij}=\frac{1-f(r)}{r^2},\quad (3\leq i<j \leq D),\nonumber\\
\tilde{R}^{34}_{~~34}&=&\left\{[\frac{6-5f(r)-f(r)^2}{r^4}+\frac{f(r)f(r)'}{2r^3}
+\frac{10-r p(r)p(r)'}{2}]\sin^2\theta\right.\nonumber\\
&-&\left.\frac{9}{r^2l^2}-\frac{6(1-f(r))}{r^4}\right\}a^2, \nonumber\\
\tilde{R}^{3k}_{~~3k}&=&2a^2\cos^2\theta(-\frac{1}{r^2l^2}
+\frac{f(r)-1}{r^4}),\quad (5\leq k \leq D) \nonumber\\
\tilde{R}^{4l}_{~~4l}&=&\left\{[\frac{f(r)f(r)'}{2r^3}
-\frac{r p(r)p(r)'}{2}+\frac{3}{r^2l^2}+\frac{-f(r)^2+2-f(r)}{r^4}]\sin^2\theta\right.\nonumber\\
&+&\left.\frac{2(f(r)-1)}{r^4}-\frac{2}{r^2l^2}\right\}a^2,\quad (5\leq l \leq D),\nonumber\\
\tilde{R}^{kl}_{~~kl}&=&a^2\cos^2\theta\frac{f(r)-1}{r^4},\quad (5\leq k < l \leq D).\nonumber
\end{eqnarray}

group (II):
\begin{eqnarray}
R^{12}_{~~34}&=&-ar^2\sin\theta\cos\theta p'(r),\quad
R^{13}_{~~24}=-\frac{a\cos\theta\sin\theta p'(r)}{2f(r)},\nonumber\\
R^{14}_{~~32}&=&-\frac{a\cos\theta p'(r)}{2f(r)\sin\theta},\quad
R^{12}_{~~24}=-\frac{ar\sin^2\theta}{2}(3p'(r)+r p''(r)),\nonumber\\
R^{1i}_{~~i4}&=&-\frac{ar\sin^2\theta}{2}p'(r),\quad (3\leq i \leq D).\nonumber
\end{eqnarray}

group (III):
\begin{eqnarray}
R^{12}_{~~13}&=&[-\frac{f'(r)}{2f(r)r^2}-p(r)^2 r-\frac{3}{r^3}
+\frac{p(r)^2f'(r)r^2}{2f(r)}-p(r)p'(r)r^2\nonumber\\
&+&\frac{3f(r)}{r^3}-\frac{f'(r)}{r^2}]a^2\sin\theta\cos\theta,\nonumber\\
R^{24}_{~~34}&=&[-p(r)p'(r)r^2+\frac{3f(r)(1-f(r))}{r^3}
+\frac{3f'(r)f(r)}{2r^2}]a^2\sin\theta\cos\theta.\nonumber
\end{eqnarray}

With regard to Ricci scalar, the term $\tilde{R}$ which is proportional to $a^2$ is given by
\begin{eqnarray}
R&=&\left\{[(\frac{r^2f(r)''}{f(r)}-8-\frac{r^2f(r)'^2}{2r^2f(r)^2}+\frac{4rf(r)'}{f(r)}
+\frac{8}{f(r)})p(r)^2+(14r-\frac{r^2f(r)'}{f(r)})p(r)'\right. \nonumber\\
&-&\left. \frac{3r^2p(r)'^2}{2}-\frac{2f(r)'}{f(r)r^3}-\frac{2f(r)f(r)'}{r^3}+\frac{f(r)'^2}{r^2}
-\frac{8}{r^4f(r)}+\frac{f(r)f(r)''}{r^2}-\frac{2f(r)'}{r^3}\right. \nonumber\\
&+&\left.\frac{f(r)'^2}{2f(r)^2r^2}-\frac{f(r)''}{r^2}-\frac{2f(r)^2}{r^4}
+\frac{20}{r^4}-\frac{f(r)''}{f(r)r^2}-\frac{10f(r)}{r^4}]\sin^2\theta
\frac{4p(r)^2}{f(r)}-\frac{14}{r^4}+ \frac{4}{f(r)r^4} \right. \nonumber\\
&+&\left.\frac{6f(r)'}{r^3}+\frac{f(r)''}{r^2}
+\frac{10f(r)}{r^4}+\frac{30\sin^2\theta-26}{r^2l^2}\right\}a^2.\nonumber
\end{eqnarray}

{\bf Acknowledgment }
This work has been supported by the Natural Science Foundation of China
under grant Nos.10875060, 10975180 and 11047025.



\begin{thebibliography}{99}
\itemsep=-4pt plus.2pt minus.2pt  
\small
\bibitem{Myers:1988ze}
Myers R C and Simon J Z
  1988 {\textit Phys. Rev.}  D {\bf 38} 2434.
\bibitem{Stelle:1976gc}
  Stelle K S
  1977 {\textit Phys. Rev.}  D {\bf 16} 953.
\bibitem{deBerredoPeixoto:2003pj}
   Berredo-Peixoto G de and Shapiro I L
  2004 {\textit Phys. Rev.}  D {\bf 70} 044024.
\bibitem{deBerredoPeixoto:2004if}
  Berredo-Peixoto G de and Shapiro I L
 2005 {\textit Phys. Rev.}  D  {\bf 71} 064005.
\bibitem{Lust:1989tj}
  Lust D and Theisen S
  1989 {\textit Lect. Notes Phys.}  {\bf 346} 1.
\bibitem{Myers:1987yn}
  Myers R C
  1987 {\textit Phys. Rev.}  D {\bf 36} 392.
\bibitem{Alekseev:1997wy}
  Alekseev S O and Pomazanov M V
  1997 {\textit Grav. Cosmol.}  {\bf 3} 191.
\bibitem{Maeda:2009uy}
  Maeda K i, Ohta N and Sasagawa Y
  2009 {\textit Phys. Rev.}  D {\bf 80} 104032.
\bibitem{Lovelock:1971yv}
  Lovelock D
  1971 {\textit J. Math. Phys.}  {\bf 12} 498.
\bibitem{Zwiebach:1985uq}
  Zwiebach B
  1985 {\textit Phys. Lett.}  B {\bf 156} 315.
\bibitem{Zwiebach:1986uq}
  Zumino B
  1986 {\textit Phys. Rept.}  {\bf 137} 109.
\bibitem{Boulware:1985wk}
  Boulware D G and Deser S
  1985 {\textit Phys. Rev. Lett.}  {\bf 55} 2656.
\bibitem{Wheeler:1985qd}
  Wheeler J T
  1986 {\textit Nucl. Phys.}  B {\bf 273} 732.
\bibitem{Cai:2003gr}
  Cai R G and Guo Q
  2004 {\textit Phys. Rev.} D {\bf 69} 104025.
\bibitem{Cai:2001dz}
  Cai R G
  2002 {\textit Phys. Rev.}  D {\bf 65} 084014
\bibitem{Cho:2002hq}
  Cho Y M and Neupane I P
  2002 {\textit Phys. Rev.}  D {\bf 66} 024044.
 \bibitem{Cvetic:2001bk}
  Cvetic M, Nojiri S and Odintsov S D
  2002 {\textit Nucl. Phys.}  B {\bf 628} 295.
\bibitem{Chen:2009an}
  Chen J and Wang Y
  2010 {\textit Chin. Phys.}  B {\bf 19} 060401.
\bibitem{Liu:2007zze}
  Liu J
  2007 {\textit Commun. Theor. Phys.}  {\bf 47} 647.
\bibitem{Dehghani:2009zzb}
  Dehghani M H and Pourhasan R
  2009 {\textit Phys. Rev.} D {\bf 79} 064015.
\bibitem{Ge:2009ac}
  Ge X H, Sin S J, Wu S F and Yang G H
  2009 {\textit Phys. Rev.} D {\bf 80} 104019.
\bibitem{Zou}
   Zou D C, Yue R H and Yang Z Y
   2011 {\textit Commun. Theor. Phys.}  {\bf 55} 449.
\bibitem{Zou:2010dx}
  Zou D C, Yang Z Y and Yue R H
  2011 {\textit Chin. Phys. Lett.}  {\bf 28} 020402.
\bibitem{Aliev:2007qi}
   Aliev A N
  2007 {\textit Phys. Rev.}  D {\bf 75} 084041.
\bibitem{Aliev:2006yk}
  Aliev A N
  2006 {\textit Phys. Rev.}  D {\bf 74} 024011.
\bibitem{Komar:1958wp}
  Komar A
 1959 {\textit Phys. Rev.}  {\bf 113} 934.
\bibitem{Peng:2007pk}
  Peng J J and Wu S Q
  2008 {\textit Chin. Phys.}  B {\bf 17} 825.
\bibitem{Zeng:2008zza}
  Zeng X X, Yang S Z and Chen D Y
  2008 {\textit Chin. Phys.}  B {\bf 17} 1629.
\end{thebibliography}
\end{document}